\documentclass[10pt,aps,prd,twocolumn,nofootinbib,superscriptaddress]{revtex4-2}
\usepackage[utf8]{inputenc}
\usepackage{color}
\usepackage{graphicx}
\usepackage{amsmath}
\usepackage{amssymb}
\usepackage{bm}
\usepackage{acronym}
\usepackage{ifthen}
\usepackage{blindtext}
\usepackage[normalem]{ulem}
\usepackage{hyperref}
\usepackage{etoolbox}
\usepackage{fancyhdr}
\usepackage{xspace}
\usepackage{textcomp}
\usepackage{multirow}
\usepackage[caption=false]{subfig}
\usepackage{tabularx}
\usepackage{soul}

\hypersetup{
    colorlinks=true,
    linkcolor=blue,
    filecolor=magenta,      
    urlcolor=black,
		citecolor=red,
		}

\newtoggle{fullauthorlist}
\toggletrue{fullauthorlist}
\newtoggle{endauthorlist}
\toggletrue{endauthorlist}

\begin{document}
\title{On Reference Frames and Coordinate Transformations}
\author{F. L. Carneiro}
\email{fernandolessa45@gmail.com}
\affiliation{Universidade Federal do Norte do Tocantins, 77824-838, Aragua\'ina, TO, Brazil, Brazil}
\author{S. C. Ulhoa}
\email{sc.ulhoa@gmail.com}
\affiliation{Instituto de F\'{\i}sica, Universidade de Bras\'{\i}lia, 70.919-970 Bras\'{\i}lia DF, Brazil}
\author{M. P. Lobo}
\email{mplobo@mail.uft.edu.br}
\affiliation{Universidade Federal do Norte do Tocantins, 77824-838, Aragua\'ina, TO, Brazil, Brazil}

\begin{abstract}
This article explores the differences between frame and coordinate transformations in relativistic theories. We highlight the key role of tetrad fields in connecting spacetime and frame indices. Using Maxwell's electrodynamics as an example, we show that Maxwell's equations are invariant under coordinate transformations but exhibit covariant behavior under frame transformations. We also analyze the energy-momentum of an electromagnetic field in different frames, providing deeper insights into the implications of different frames of reference and coordinate systems.
\end{abstract}

\maketitle
\section{Introduction}

Contemporary natural philosophy describes observed phenomena in nature using the usual mathematical language. From a physical perspective, the motion of a body is considered the most elementary classical phenomenon. This motion can be quantitatively described using a ruler or a protractor, combined with a clock, for measuring distance or angle and time, respectively. To translate these measurements into an interpretation of the phenomenon, one must compare the numbers to a standard position, known as the origin, and a standard time. Once the origin is established, one can construct a coordinate system containing three linearly independent axes. A coordinate system translates measured values into physical information, e.g., if the ruler's position changes over time, an observer may conclude that they are measuring motion. However, another observer might disagree with the first; they might measure motion at a different rate (velocity) or may not perceive any motion at all, even when using the same coordinate system.

The previous example illustrates that a coordinate system alone does not provide enough information to fully characterize a physical phenomenon; thus, establishing a frame of reference is necessary. On the one hand, a coordinate system represents a passive transformation; on the other hand, a reference system represents an active transformation. An observer employs a coordinate system and carries along a reference frame adapted to themselves throughout their trajectory. Ergo, one needs to specify both the coordinate system used for measurement and the frame of reference. For instance, the velocity $v$ of a body is measured using Cartesian coordinates in a frame of reference denoted as $S$. 

While we can merge the concepts of the observer and the frame, referring to the observer as the frame, it is not feasible to mix the concepts of coordinate and frame. Coordinates are merely a mathematical tool—a language used to describe phenomena. The choice of coordinates does not alter the qualitative description of a physical phenomenon. For instance, the velocity of a body can be measured using Cartesian, spherical, cylindrical, or other coordinate systems, yet the existence of motion is independent of the chosen coordinates. The distinct natures of coordinates (mathematical) and frames (physical) underscore that they should be treated as separate entities and, most importantly, transformed independently.

The rules governing frame transformations emerge from the foundational assumptions of a theory. Assuming that the time interval and the length of a body remain invariant during frame transformations leads us to the Galilean transformation
\begin{equation}\label{eq1}
\left(
\begin{array}{c}\tilde{t}\\\tilde{x}\\\tilde{y}\\\tilde{z}
\end{array}
\right)=
\left(
\begin{array}{cccc}1&0&0&0\\
-v&1&0&0\\
0&0&1&0\\
0&0&0&1
\end{array}
\right)\left(
\begin{array}{c}t\\x\\y\\z
\end{array}
\right)\,,
\end{equation}
where $v$ represents the relative velocity of the frames, with the assumption of relative motion along the $x$ axis. Consequently, the position $x^{(i)}$ as measured by the frame $S$ can be compared with $\tilde{x}^{(i)}$, as measured in frame $\tilde{S}$. Newtonian Mechanics (NM) inevitably arises from adopting these non-symmetrical transformations (\ref{eq1}). These transformations can be easily extended to other spatial coordinates. However, this extension is beyond the scope of this article, as the distinction between physical and mathematical transformations is sufficiently demonstrated in the simplest case.

In textbooks, the Galilean transformations are typically obtained by constructing two coordinate systems: one $\big(t,x^{i}\big)$ specifically adapted to $S$ and another $\big(t',x'^{i}\big)$ correspondingly adapted to $\tilde{S}$. The transformation rules are derived by relating the origins of the two coordinate systems, specifically, $x' = x - vt$. Albeit the standard textbook method produces a transformation between frames, assuming that $\tilde{S}$ carries its coordinate system along the trajectory, as in $\big(\tilde{t}, \tilde{x}^{(i)}\big) = \big(t', x'^{i}\big)$, this procedure actually does not represent a frame transformation.

If the requirements of invariant time and length are relaxed, and instead, we assume the velocity of information propagation as the invariant, we arrive at a new transformation rule for the frames, specifically, the symmetrical Lorentz transformations
\begin{equation}\label{eq2}
\left(
\begin{array}{c}\tilde{t}\\\tilde{x}\\\tilde{y}\\\tilde{z}
\end{array}
\right)=
\left(
\begin{array}{cccc}\gamma&-\beta\gamma&0&0\\
-\beta\gamma&\gamma&0&0\\
\,\,0&\,\,0&\,\,1&\,\,0\\
\,\,0&\,\,0&\,\,0&\,\,1
\end{array}
\right)\left(
\begin{array}{c}t\\x\\y\\z
\end{array}
\right)\,,
\end{equation}
where $\beta \equiv v/c$, $\gamma=\big(1-\beta^{2}\big)^{-1/2}$ with $c$ being the speed of light. The same discussion applicable to transformation rule (\ref{eq1}) also applies to (\ref{eq2}), indicating that the transformation between frames can be achieved through a coordinate transformation between two coordinate systems tailored to each observer. The transformation is not merely a translation of the origin but represents a four-dimensional rotation along the $t,x$ plane. Such a procedure is adopted in several textbooks \cite{brau2004modern,rindler1977essential,ferraro2007einstein,gazzinelli2019teoria}.The intrinsic relation between time and spatial quantities in (\ref{eq2}) highlights the necessity of constructing a pseudo-Euclidean, fictional, four-dimensional space, called spacetime \cite{landau2013classical}. Upon adopting transformation rule (\ref{eq2}), one arrives at Relativistic Mechanics (RM).

In both theories, we can alter coordinates in the same way, since coordinate transformation represents not a physical reality but a mathematical procedure. However, altering the method of changing frames necessitates a shift in theories. Although it may appear trivial, the rigor in distinguishing between coordinate and frame transformations in NM and RM becomes critical. Problems arise when these same procedures are applied in General Relativity (GR), particularly when the restriction of inertial frames is relaxed—for example, when considering accelerated frames in relativistic theories like Maxwell's Electrodynamics (ME).

Since the birth of GR, modern gravitational theories have shifted from describing the interactions between massive bodies by means of a gravitational force to employing geometrical descriptions of gravity. When constructing a geometrical theory of gravity, we must choose a geometry in which to establish the theory. GR is based on Riemannian geometry, within which the existence of gravitational effects in spacetime is characterized by the presence of non-trivial Riemannian geometry, namely, a non-vanishing curvature tensor. Thus, the presence of the curvature tensor directly indicates the presence of a gravitational field. The geometrical description of spacetime is achieved by means of the square infinitesimal distance (interval) between two neighboring points. To evaluate such a distance, localizing these points with a coordinate system is necessary, meaning a choice of coordinate system is crucial for characterizing the geometry. Computing two distinct intervals may involve either the same spacetime described in two different coordinate systems or two entirely distinct spacetimes. Consequently, coordinates hold great significance in GR, and confusing coordinates with frame transformations can result in numerous misconceptions.

In this paper, we aim to clearly and formally distinguish between coordinates and frame transformations in physical theories, a task presently unaddressed in the literature to our knowledge. We begin in section \ref{coordinates} by examining how tensors can be transformed between coordinate systems and how tensors are defined from their transformation rule. Additionally, we present the generalization of differential operators for  an arbitrary coordinate system. In section \ref{frames}, we distinguish between spacetime and frame indices and introduce the tetrad fields as projectors of spacetime quantities into frames. With the distinction between frames and coordinate transformations firmly established, in section \ref{maxwell}, we illustrate this distinction within ME. We demonstrate the covariance of the field equations of the theory when switching between frames without resorting to coordinate transformations. Subsequent to the analysis, we detail how to properly project the energy-momentum tensor of the electromagnetic field into a frame. Finally, in section \ref{conclusions}, we offer our conclusions.

Throughout this article, we employ the following notation. We denote spacetime indices with Greek letters $\mu, \nu, \ldots$, ranging from $0$ to $3$. Frame indices are indicated by Latin letters from the beginning of the alphabet, such as $a, b, \ldots$, and they range from $(0)$ to $(3)$. When two identical contracted indices appear, it signifies summation, e.g., $x_{\mu}x^{\mu}\equiv \displaystyle{\sum_{\mu=0}^{3}}x_{\mu}x^{\mu}$. We use primes to denote distinct coordinates and tildes to distinguish between different frames.

\section{Coordinate transformations}\label{coordinates}

Physical laws represent relations between physical quantities. In general, the positions of particles in time and space are chosen as the fundamental quantities. We can write Newton's Second Law in a specific basis $\{\hat{e}_{1},\hat{e}_{2},\hat{e}_{3}\}$ as
\begin{equation}\label{eq3a}
\vec{F}=m \big( a^{1}\hat{e}_{1} + a^{2}\hat{e}_{2} + a^{3}\hat{e}_{3} \big) = m \, a^{i}\hat{e}_{i}\,,
\end{equation}
where $i=1,2,3$. In Cartesian coordinates, the basis are $\{\hat{e}_{i}\}=\{\hat{x},\hat{y},\hat{z}\}$ and we may write (\ref{eq3a}) as
\begin{equation}\label{eq3b}
\vec{F}=m \big( a_{x}\hat{x} + a_{y}\hat{y} + a_{z}\hat{z} \big)\,,
\end{equation}
where $a_{x,y,z}=a^{1,2,3}$\,. In cylindrical coordinates, the basis are $\{\hat{e}_{i}\}=\{\hat{\rho},\hat{\phi},\hat{z}\}$ and we may write (\ref{eq3a}) as
\begin{equation}\label{eq3c}
\vec{F}=m \big( a_{\rho}\hat{\rho} + a_{\phi}\hat{\phi} + a_{z}\hat{z} \big)\,.
\end{equation}
While the components in the two coordinate systems differ, the underlying equations of motion remain the same. Let us consider an example of a motion where $a_{\rho}=1$ and $a_{\phi}=0=a_{z}$. From (\ref{eq3c}), we have
\begin{equation}\label{eq3d}
\vec{F}=m \hat{\rho}\,.
\end{equation}
Given that  $\hat{\rho}=\cos{\phi}\hat{x}+\sin{\phi}\hat{y}$, it follows that
\begin{equation}\label{eq3e}
\vec{F}=m \big( \cos{\phi}\hat{x}+\sin{\phi}\hat{y} \big)\,.
\end{equation}
By identifying $\cos{\phi}=a_{x}$ and $\sin{\phi}=a_{y}$, we recover (\ref{eq3b})\,.

We may write Newton's Second Law in many coordinate systems, but in all cases, the physical behavior remains unchanged, that is, physical laws are invariant under coordinate transformations. We can eliminate the dependence of the coordinate system in our notation by expressing it in a vector form as
\begin{equation}\label{eq3}
\vec{F}=m\vec{a}\,.
\end{equation}
The vector form (\ref{eq3}) is unique, since one does not need to specify the coordinate system. This observation occurs because vectors are geometrical entities existing independently of any coordinate system. Ergo, it is convenient to express physical laws by means of vectors and other tensors. The process of attributing coordinates to physical reality in order to describe it defines a mathematical quantity known as Manifold. In NM, the manifold is represented by a three-dimensional Euclidean space. Thus, utilizing vectors in the description of physical laws extends their applicability to any coordinate system within this manifold.


The disentangled nature of NM between space and time allows the construction of three-dimensional vectors $\vec{v}$ to describe the observed physical quantities. However, due to the relativistic nature of RM and ME, it is necessary to consider transformations in both time and space. Therefore, adopting Minkowski's approach, which utilizes a non-Euclidean manifold, enables the expression of relativistic quantities by means of 4-vectors $\bf{v}$, where the $\mu=0$ component represents a distance derived from time multiplied by $c$, while the components $\mu=i=1,2,3$ correspond to the usual spatial ones.

We may represent a four-vector in a basis $\bf{\hat{e}}$ as 
\begin{equation}\label{eq4}
\textbf{v}=v^{\mu}\hat{e}_{\mu}=v_{\mu}\hat{e}^{\mu}\,,
\end{equation}
where $v^{\mu}$ and $v_{\mu}$ are distinct representations of the components of $\bf{v}$. In what follows, we shall refer to the components of the vectors as the vector itself. We can separate the temporal and spacial components as
\begin{equation}\nonumber
v^{\mu}=\{v^{0},v^{i}\}=\{v^{0},v^{1},v^{2},v^{3}\}\,.
\end{equation}

In order to properly define 4-vectors (or any vector of any dimension) we shall consider the transformation rule of its components between distinct basis.
Let us consider two coordinate systems $x^{\mu}$ and $x'^{\mu}$.
Under coordinate transformations, the quantities that transform according to
\begin{equation}\label{eq5}
v'^{\mu}=\frac{\partial x'^{\mu}}{\partial x^{\nu}}v^{\nu}
\end{equation}
are known as contravariant components, while those that transform as
\begin{equation}\label{eq6}
v'_{\mu}=\frac{\partial x^{\nu}}{\partial x'^{\mu}}v_{\nu}
\end{equation}
are covariant components.
The transformation rules (\ref{eq5}) and (\ref{eq6}) can be used to define four-vectors, i.e., a four-vector $v^{\mu}$ is defined as something that transforms according to (\ref{eq5}). 

By definition, scalar quantities have no indices. Therefore, we can define a scalar $\Phi$ as a quantity that transforms between two coordinate systems according to
\begin{equation}\nonumber
\Phi'(x'^{\mu})=\Phi(x^{\mu})\,.
\end{equation}
This means they are invariant under coordinate transformations.

Vectors are tensors of rank 1, just as scalars are tensors of rank 0. We can generalize the transformation rule (\ref{eq5}) and define tensors of rank 2, $T^{\mu\nu}$, which are entities that transform as
\begin{equation}\label{eq7}
T'^{\mu\nu}=\frac{\partial x'^{\mu}}{\partial x^{\alpha}}\frac{\partial x'^{\nu}}{\partial x^{\beta}}T^{\alpha\beta}\,.
\end{equation}
Following this procedure, we define the transformation rule for an arbitrary tensor as
\begin{equation}\label{eq8}
\hspace{-0.2cm}T'^{\mu\ldots\nu}\,_{\alpha\ldots\beta}=\frac{\partial x'^{\mu}}{\partial x^{\rho}}\cdots\frac{\partial x'^{\nu}}{\partial x^{\lambda}}\frac{\partial x^{\gamma}}{\partial x'^{\alpha}}\cdots\frac{\partial x^{\delta}}{\partial x'^{\beta}}T^{\rho\ldots\lambda}\,_{\gamma\ldots\delta}\,.
\end{equation}
It is important to note that while a tensor of rank $N$ possesses $N$ indices, not all quantities with indices are tensors. To be considered a tensor, a quantity must transform as shown in equation (\ref{eq8}). When the indices of a tensor can be permuted without changing its value, as in $T^{\mu\nu}=T^{\nu\mu}$, the tensor is said to be symmetrical; if permuting the indices results in the opposite value, as in $T^{\mu\nu}=-T^{\nu\mu}$, it is termed antisymmetric.

Tensors, being geometric entities, exist independently of coordinates. When we describe measurable physical quantities as tensors and formulate physical laws in a tensorial framework, they achieve invariance under coordinate transformations, rendering them ideal for describing natural phenomena. A wide number of tensors hold significant roles in the realm of physics. These include tensors that represent fundamental physical quantities, such as the stress-energy tensor, the inertia tensor, and the Faraday tensor. Others, such as the Kronecker delta\footnote[1]{{The Kronecker delta is represented by the identity matrix, denoted as $\delta^{\mu}_{\nu}$, with $\delta^{\mu}_{\nu}=diag(1,1,1,1)$ in four dimensions. It yields a value of $1$ if $\mu=\nu$ and $0$ if $\mu\neq\nu$. Consequently, $\delta^{\mu}_{\nu}A_{\mu}$ simplifies to $A_{\nu}$.}} $\delta^{\mu}_{\nu}$ and the Levi-Civita\footnote[2]{The Levi-Civita pseudo-tensor is entirely antisymmetric. It yields a value of $1$ if $\alpha\beta\mu\nu=0,1,2,3$ or any even permutation, $-1$ for an odd permutation, and $0$ if at least two indices repeat.} $\epsilon^{\alpha\beta\mu\nu}$ pseudo-tensor\footnote[3]{A pseudo-tensor behaves as a tensor in transformations but yields a negative sign in the reverse transformation.}, serve as critical tools for mathematical manipulation. Beyond their mathematical utility, tensors like the metric tensor hold not only mathematical significance but also convey valuable physical information in GR.

\subsection{The metric tensor}

In equation (\ref{eq4}), we represented the vector using both its covariant and contravariant components. The specific form chosen to represent a vector does not affect its inherent properties, but having the capability to transform between these forms is crucial.

There are two ways to represent the square of $\bf{v}$. First, we can evaluate the sum of the squares of its components
\begin{equation}\label{eq9}
\textbf{v}\cdot\textbf{v}=v^{\mu}v_{\mu}\,.
\end{equation}
Second, we can use (\ref{eq4}) as
\begin{equation}\label{eq10}
\textbf{v}\cdot\textbf{v}=v^{\mu}v^{\nu}\hat{e}_{\mu}\hat{e}_{\nu}\,.
\end{equation}
By comparing (\ref{eq9}) and (\ref{eq10}), we can see that
\begin{equation}\nonumber
v_{\mu}=g_{\mu\nu}v^{\nu}\,,
\end{equation}
where we define the tensor $g_{\mu\nu}\equiv \hat{e}_{\mu}\hat{e}_{\nu}$. The symmetric tensor $g_{\mu\nu}$ is known as the metric tensor. This tensor possesses ten independent components and can be employed to raise and lower indices in other tensors. For example, $T^{\mu}\,_{\nu}\,^{\lambda}=g^{\mu\alpha}g_{\nu\beta}T_{\alpha}\,^{\beta\lambda}$.

The spacetime interval, denoted as $ds^2$, is a cornerstone of the theory of relativity, serving as a measure of separation between two points in spacetime. However, the exact form of this interval varies depending on the geometry of the spacetime being considered. The expression $ds^2=dx^{\mu}dx_{\mu}$ represents a general formulation applicable in various physical contexts.
The introduction of the metric tensor enables us to raise the second index of $ds^2$, thus resulting in
\begin{equation}\label{eq11}
ds^{2}=g_{\mu\nu}dx^{\mu}dx^{\nu}\,.
\end{equation}
The metric tensor is more than a mathematical construct; it offers crucial insights into the structure of spacetime within a given coordinate system. In the Cartesian coordinates of Minkowski spacetime, the metric tensor assumes the diagonal form $g_{\mu\nu}=diag(-1,1,1,1)$.

\subsection{Tensorial calculus}

By choosing to represent physical quantities as tensors, one embraces a formalism that is independent from any specific coordinate system. This mathematical approach is in harmony with the fundamental principle of relativistic physics, which dictates that physical laws should remain consistent regardless of the chosen coordinates. Tensors provide us with this desired invariance under coordinate transformations.

However, physical laws often involve the differentiation of tensors. In the pursuit of a consistent and universally applicable framework, these derivatives must also be tensors. In other words, the fundamental requirement is that the derivatives of tensors, representing the rates of change of physical quantities, should transform covariantly under coordinate transformations. This covariant transformation property ensures that the mathematical representation of physical laws remains consistent and meaningful across various coordinate systems, thereby reflecting the underlying physical reality and its inherent invariance.

Considering a rank 1 tensor $A^{\mu}$, which is a four-vector, its derivative transforms as
\begin{small}
\begin{equation}\label{eq12}
\partial'_{\nu}A'^{\mu}\equiv\frac{\partial A'^{\mu}}{\partial x'^{\nu}}=\frac{\partial x'^{\mu}}{\partial x^{\alpha}}\frac{\partial x^{\beta}}{\partial x'^{\nu}}\frac{\partial A^{\alpha}}{\partial x^{\beta}}+\frac{\partial^{2}x'^{\mu}}{\partial x^{\alpha}\partial x^{\beta}}\frac{\partial x^{\beta}}{\partial x'^{\nu}}A^{\alpha}\,.
\end{equation}
\end{small}
The first term on the RHS of (\ref{eq12}) behaves as expected for a tensor transformation. However, the presence of the second term on the RHS indicates that the derivative of a tensor is not a tensor. Therefore, we must define another derivative by introducing a non-tensor term $\Gamma^{\lambda}\,_{\mu\nu}$ that transforms as
\begin{small}
\begin{equation}\label{eq13}
\Gamma'^{\lambda}\,_{\mu\nu}=\frac{\partial x^{\alpha}}{\partial x'^{\mu}}\frac{\partial x^{\beta}}{\partial x'^{\nu}}\frac{\partial x'^{\lambda}}{\partial x^{\gamma}}\Gamma^{\gamma}\,_{\alpha\beta}-\frac{\partial^{2}x'^{\lambda}}{\partial x^{\alpha}\partial x^{\beta}}\frac{\partial x^{\alpha}}{\partial x'^{\mu}}\frac{\partial x^{\beta}}{\partial x'^{\nu}}\,.
\end{equation}
\end{small}
With the aid of definition (\ref{eq13}), we can define the covariant derivative of contravariant components as
\begin{equation}\nonumber
\nabla_{\nu}A^{\mu}\equiv\partial_{\nu}A^{\mu}+\Gamma^{\mu}\,_{\nu\lambda}A^{\lambda}
\end{equation}
and of the covariant components as
\begin{equation}\nonumber
\nabla_{\nu}A_{\mu}\equiv\partial_{\nu}A_{\mu}-\Gamma^{\lambda}\,_{\mu\nu}A_{\lambda}\,.
\end{equation}
The non-tensorial quantity that transforms as described in (\ref{eq13}) is known as the affine connection. The affine connection plays a crucial role in differential geometry and provides a formalism for understanding how the components of vectors change when we transport them through space.
In the realm of differential geometry, it's important to recognize that vectors exist in a manner independent of the coordinate system in which they are expressed. Vectors are mathematical entities with components that can be defined in various coordinate systems, including Cartesian, polar, and others. The transformation of vectors between these coordinate systems is governed by the affine connection, which captures the variation of vector components with respect to changes in the coordinate basis.
For instance, consider a plane, and imagine transporting a vector between two points. In a Cartesian coordinate system, the components of the vector remain unaltered during this process. This is a reflection of the fact that Cartesian basis vectors retain their direction and do not change as we move through space. In contrast, when working in a polar coordinate system, the situation is different. The polar basis vectors change direction relative to the Cartesian ones as we transport the vector. Consequently, the components of the vector in polar coordinates differ from those in Cartesian coordinates due to this change in basis vectors' orientation.

The covariant derivative can also be expressed as a directional derivative, along the tangent vector $u^{\nu}=dx^{\nu}/d\tau$ field, as
\begin{equation}\nonumber
\frac{D\, A^{\mu}}{d\tau}=u^{\nu}\nabla_{\nu}A^{\mu}\,,
\end{equation}
where $\tau$ is an affine parameter\footnote[4]{An affine parameter is a quantity that allows for the parametrization or description of a curve or trajectory in a particular manner. It can be a measure of distance, time, or any other quantity that varies along the trajectory. The term ``affine"} suggests that this quantity can be expressed linearly or proportionally along the curve, and its selection can simplify the study of the curve's properties., e.g., the proper time of the observer or any linear transformation of it.

We can generalize to tensors of higher order. For each contravariant component of the tensor, an affine connection is added, and for each covariant component, a connection is subtracted. For example, in the case of a tensor of rank 2,
\begin{equation}\label{eq14}
\nabla_{\lambda}T^{\mu\nu}=\partial_{\lambda}T^{\mu\nu}+\Gamma^{\mu}\,_{\gamma\lambda}T^{\gamma\nu}+\Gamma^{\nu}\,_{\gamma\lambda}T^{\mu\gamma}\,.
\end{equation}
When we permute the lower indices of the connection and subtract one from another, we obtain the torsion tensor
\begin{equation}\nonumber
T^{\lambda}\,_{\mu\nu}\equiv\Gamma^{\lambda}\,_{\mu\nu}-\Gamma^{\lambda}\,_{\nu\mu}
\end{equation}
of the manifold.

When developing a geometric theory, a fundamental choice involves determining the geometry of spacetime. If we choose a manifold metricity, i.e.,
\begin{equation}\label{eq15}
\nabla_{\lambda}g_{\mu\nu}=\partial_{\lambda}g_{\mu\nu}+\Gamma^{\gamma}\,_{\mu\lambda}g_{\gamma\nu}+\Gamma^{\gamma}\,_{\nu\lambda}g_{\mu\gamma}=0,
\end{equation}
we can cyclically permute the indices in (\ref{eq15}), combine the three resulting equations, and express the Christoffel symbols as
\begin{equation}\label{eq16}
\Gamma^{\lambda}\,_{\mu\nu}=\mathring{\Gamma}^{\lambda}\,_{\mu\nu}-\frac{1}{2}\left(T_{\mu\nu}\,^{\lambda}+T_{\nu}\,^{\lambda}\,_{\mu}+T^{\lambda}\,_{\nu\mu}\right)\,.
\end{equation}
If we choose a geometry with a zero torsion tensor, i.e., $T^{\lambda}\,_{\mu\nu}=0$, the affine connection can be expressed as
\begin{equation}\label{eq17}
\mathring{\Gamma}^{\lambda}\,_{\mu\nu}=\frac{1}{2}g^{\lambda\gamma}\left(\partial_{\mu}g_{\gamma\nu}+\partial_{\nu}g_{\nu\gamma}-\partial_{\gamma}g_{\mu\nu}\right)\,,
\end{equation}
where the components of the affine connection, denoted as $\mathring{\Gamma}^{\lambda}\,_{\mu\nu}$, hold particular significance and are commonly referred to as Christoffel symbols.

A noteworthy point to consider is that when we have a manifold characterized by metricity, meaning the condition $\nabla_{\lambda}g_{\mu\nu}=0$, and a vanishing torsion tensor, i.e., $T^{\lambda}\,_{\mu\nu}=0$, we are dealing with a Riemannian manifold. Riemannian manifolds, with their metric structure and absence of torsion, provide the basis for the spacetime geometry in the framework of General Relativity (GR).
This choice of geometry is fundamental in the context of gravitational physics, as it underlies Einstein's theory of gravity, where the curvature of spacetime is governed by the metric tensor and described by the equations of GR. The Christoffel symbols, derived from this geometric framework, are a key component in formulating the geodesic equations that describe the motion of particles and test bodies in gravitational fields.

\section{Frame transformations}\label{frames}

A frame is a system employed by an observer to quantify physical quantities accurately. These quantities are represented by tensors, and to attribute a physical interpretation to them, they must be projected into a specific frame. Tensors are defined within a coordinate system, and by themselves, they lack intrinsic physical significance until projected into a frame. Consequently, frames possess physical properties, whereas coordinates do not.

As we transition between different frames, it is reasonable to anticipate transformations affecting physical quantities. Thus, when measuring a physical quantity, it is imperative to specify the frame from which it was measured. This emphasis on frame specification underscores the intricate relationship between the observer's chosen frame and the accurate description of physical phenomena, emphasizing the crucial role that frames play in the realm of physics.

\begin{figure}[htbp]
	\centering
		\includegraphics[width=0.40\textwidth]{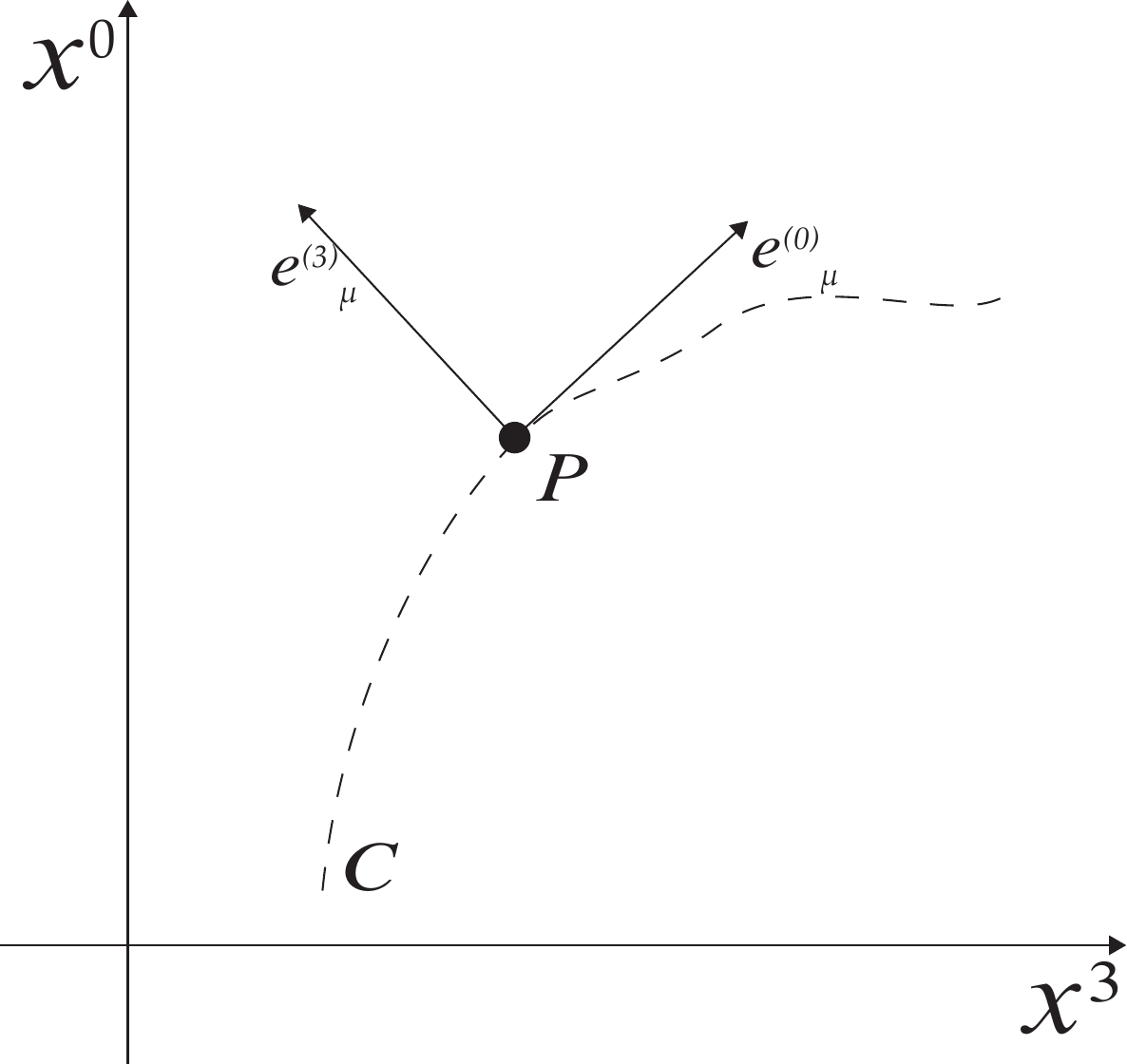}
	\caption{Instantaneous Rest Frame (IRF) at an event $P$ on the worldline $\mathcal{C}$. For visualization, the other two spatial dimensions were omitted.}
	\label{fig1}
\end{figure}

Let us consider an observer who carries with them a dedicated coordinate system in which they are always positioned at the origin. Physical quantities measured by this observer are denoted with Latin letters from the beginning of the alphabet, e.g., the observer measures the velocity as $v^{a}=\{v^{(0)},v^{(1)},v^{(2)},v^{(3)}\}$ for a particle whose four-velocity is represented by the tensor $v^{\mu}=\{v^{0},v^{1},v^{2},v^{3}\}$. When this observer measures the four-velocity of the particle, he projects it into his own frame as
\begin{equation}
v^{a}=e^{a}\,_{\mu}v^{\mu}\,,
\end{equation}
where the quantities $e^{a}\,_{\mu}$ are know as tetrads. Tetrads possess two indices: one Greek and one Latin. The Greek index transforms as a four-vector under coordinate transformations, i.e.,
\begin{equation}\label{eq19}
e'^{a}\,_{\mu}=\frac{\partial x^{\nu}}{\partial x'^{\mu}}e^{a}\,_{\nu}\,,
\end{equation}
while the Latin index does not change. The set of tetrads consists of four 4-vectors $\{ e^{(0)}\,_{\mu},e^{(1)}\,_{\mu},e^{(2)}\,_{\mu},e^{(3)}\,_{\mu} \}$ in spacetime. Collectively, these tetrads establish the Instantaneous Rest Frame (IRF) of an observer moving along a worldline $\mathcal{C}$ represented by $x^{\mu}(\tau)$ in spacetime, where $\tau$ serves as an affine parameter, as can be seen in Figure \ref{fig1}.
Observations recorded at event $P$ correspond to measurements conducted within the Instantaneous Rest Frame (IRF).

The observer remains perpetually at rest within the IRF, which implies that their 4-velocity $u^{\mu}$ must be parallel to $e_{(0)},^{\mu}$. Consequently, 
\begin{equation}\label{eq21}
u^{\mu}=c\,e_{(0)}\,^{\mu}\,.
\end{equation}
From (\ref{eq21}), we can see that if $e^{a}\,_{\mu}=\delta^{a}_{\mu}$, the 4-acceleration reads
\begin{equation}\label{eq22}
a^{\mu}=\frac{D\,u^{\mu}}{d\tau}=\frac{D\,\delta_{(0)}^{\mu}}{d\tau}=0\,,
\end{equation}
i.e., the observer (and $e^{a}\,_{\mu}$) is adapted to a frame that is at rest, as indicated by the vanishing 4-acceleration.

The spacetime index in the tetrads transforms as a 4-vector, and we can raise and lower it using the metric tensor, i.e.,
\begin{equation}\label{eq23}
e^{a}\,_{\mu}=g_{\mu\lambda}e^{a\lambda}\,.
\end{equation}
By multiplying both sides of (\ref{eq23}) by $e_{a\nu}$ and noticing that $e_{a\nu}e^{a\lambda}=\delta^{\lambda}_{\nu}$, we arrive at
\begin{equation}\label{eq24}
e^{a}\,_{\mu}e_{a\nu}=g_{\mu\nu}\,.
\end{equation}
As a result, the metric tensor can be determined through the tetrads. The tetrads, represented by $e^{a}\,_{\mu}$, carry valuable information about the spacetime's geometry in which physical phenomena occur. They also contain kinematic information about the observer who measures these physical quantities. In essence, the measurement results are the outcome of projecting spacetime quantities into physical quantities through the utilization of tetrads, underscoring the integral role tetrads play in connecting spacetime geometry and the measurements of physical phenomena.

The spacetime index in the tetrads undergoes transformation as specified in (\ref{eq19}). The frame index, however, must undergo a transformation specific to frames. When we establish a frame transformation rule, we essentially choose a set of theories. In the case of selecting the Lorentz transformation, a consequence of the Principle of Relativity, we enter into the realm of relativistic theories, such as the RM and ME. Therefore, when we confine ourselves to the Lorentz transformation, the frame index of the tetrads, along with other frame-related quantities, transforms between frames as
\begin{equation}
\tilde{e}^{a}\,_{\mu}=\Lambda^{a}\,_{b}e^{b}\,_{\mu}\,,
\end{equation}
where, for a transformation along the $x^{(1)}$ axis, the Lorentz transformation matrix $\Lambda^{a}\,_{b}$ can be expressed as
\begin{equation}
\Lambda^{a}\,_{b}=\left(
\begin{array}{cccc}\gamma&-\beta\gamma&0&0\\
-\beta\gamma&\gamma&0&0\\
\,\,0&\,\,0&\,\,1&\,\,0\\
\,\,0&\,\,0&\,\,0&\,\,1
\end{array}\right)\,.
\end{equation}
Hence, for a 4-vector, we transform frames as
\begin{equation}
\tilde{v}^{a}=\Lambda^{a}\,_{b}v^{b}\,,
\end{equation}
and for coordinate transformations, we use the equation (\ref{eq5}).

The clear distinction and autonomy of frame transformations and coordinate transformations enable us to perform these transformations independently. To illustrate, let us consider a rank 2 tensor with one index projected into a frame, i.e., $T^{a\mu}$. We can independently transform the frame  $\tilde{T}^{a\mu}=\Lambda^{a}\,_{b}T^{b\mu}$, the coordinates $T'^{a\mu}=\frac{\partial x^{\mu}}{\partial x^{\nu}}T^{a\nu}$, or both simultaneously $\tilde{T}'^{a\mu}=\frac{\partial x^{\mu}}{\partial x^{\nu}}\Lambda^{a}\,_{b}T^{b\nu}$. This independence allows us to address various scenarios effectively. For instance, it allows us to study electromagnetic radiation emitted by an accelerated source in two different frames within the same coordinate system, as explored in reference \cite{maluf2010electrodynamics}. Similarly, it facilitates the analysis of gravitational energy flux in the context of a propagating accelerated gravitational bubble from both a co-moving and a static frame, all within the same coordinate system, as examined in reference \cite{carneiro2022total}.

As a simple example, let us explore two ways in which an observer measures the velocity $\tilde{v}^{a}$, represented by the 4-vector $v^{\mu}$, of a particle. The first method is:
\begin{enumerate}
	\item Project $v^{\mu}$ into the IRF of the particle: $v^{b}=e^{b}\,_{\mu}v^{\mu}$\,,
	\item Transform from the particle's frame into the observer's frame: $\tilde{v}^{a}=\Lambda^{a}\,_{b}v^{b}$\,.
\end{enumerate}
The other possibility is
\begin{enumerate}
	\item Transform the tetrads: $\tilde{v}^{a}\,_{\mu}=\Lambda^{a}\,_{b}e^{b}\,_{\mu}$\,,
	\item Project the 4-velocity directly into the frame of the observer: $\tilde{v}^{a}=\tilde{e}^{a}\,_{\mu}v^{\mu}$\,.
\end{enumerate}
In both cases, the coordinates remain unchanged. In the next section, we will explore a more complex example using the physical quantities of (ME).

\section{Maxwell's theory in an arbitrary inertial frame}\label{maxwell}

Maxwell's electrodynamics serves as the framework through which we comprehend the interaction between electric charges and the electromagnetic field. In this context, the actions of a charge $q_{1}$ on another charge $q_{2}$ manifest through the presence of an electric field $\vec{E}$ or a magnetic field $\vec{B}$, depending on the frame of reference. Classically, this interplay can be thought as being initiated by the source charge $q_{1}$, which generates a field that subsequently exerts its influence on the test charge $q_{2}$.

The classical electromagnetic field can be described mathematically in terms of two non-physical potentials: the scalar potential $\phi$ and the vector potential $\vec{A}$. These potentials are fundamental components of the electromagnetic field and are indispensable in elucidating the dynamics and interactions of electric charges with the field. They serve as intermediaries between the source charges and the resultant electromagnetic effects and can be written as
\begin{align}
\vec{E}=&-\vec{\nabla}\Phi-\partial_{t}\vec{A}\,,\nonumber\\
\vec{B}=& \vec{\nabla}\times \vec{A}\,.\nonumber
\end{align}
In any given frame, the source of the electromagnetic field can be accurately represented as comprising two essential components: a charge density $\rho$ and a surface current density $\vec{J}$. The charge density $\rho$ conveys the distribution of electric charge within the source, while the surface current density $\vec{J}$ delineates the flow of electrical currents along the source's surface.

From the scalar potential $\phi$ and the vector potential $\vec{A}$, we can construct an entity known as the 4-potential. This 4-potential, denoted as $A^{\mu}$, encapsulates both the temporal and spatial aspects of the potentials in a unified framework. By combining these potentials into a single 4-potential, we describe the electromagnetic phenomena in tensorial equations. The 4-potential is defined as
\begin{equation}
A^{\mu}\equiv \big( \Phi/c,\vec{A} \big)\,.
\end{equation}
Additionally, we can define a 4-current density, denoted as $j^{\mu}$, which complements the 4-potential. The 4-current density is a mathematical construct that describes the flow of electric charge in both the temporal and spatial dimensions of spacetime. The 4-current is defined as
\begin{equation}
j^{\mu}= \big( \rho\, c, \vec{J} \big)\,.
\end{equation}
From the 4-potential, we can derive a rank-2 antisymmetric tensor, denoted as $F^{\mu\nu}$. This tensor captures the essence of the electromagnetic field by encapsulating both the electric and magnetic field components in a compact and elegant mathematical form. The antisymmetry of this tensor is a fundamental property that reflects the intrinsic relationship between electric and magnetic fields in electromagnetism. This tensor is called Faraday tensor and can be represented as
\begin{equation}
F_{\mu\nu}\equiv \nabla_{\mu}A_{\nu} - \nabla_{\nu}A_{\mu}\,.
\end{equation}
In Cartesian coordinates, the Faraday tensor is given by
\begin{equation}\label{eq30}
F_{\mu\nu}
=\left(
\begin{array}{cccc}0&E_{x}/c&E_{y}/c&E_{z}/c\\
-E_{x}/c&0&-B_{z}&B_{y}\\
-E_{y}/c&B_{z}&0&-B_{x}\\
-E_{z}/c&-B_{y}&B_{x}&0
\end{array}
\right)\,.
\end{equation}
From the Faraday tensor, we define the dual tensor, also known as the electromagnetic dual or Hodge dual tensor. The dual tensor is a fundamental mathematical construct that encodes the duality between electric and magnetic fields in electromagnetism. It can be obtained from
\begin{equation}\label{eq31}
\mathcal{F}^{\mu\nu}=\frac{1}{2}\epsilon^{\mu\nu\alpha\beta}F_{\alpha\beta}\,,
\end{equation}
where $\epsilon^{\mu\nu\alpha\beta}$ is the Levi-Civita pseudo-tensor.

In a given frame, the electric and magnetic fields can be determined by projecting the Faraday tensor into that frame. In order to illustrate this concept, consider a frame denoted as $S$ that exclusively measures an electric field component along the $x^{(2)}$ direction. In this frame, we can express the electric field as
\begin{equation}
E_{y}=c\,F_{(0)(2)}=c\,e_{(0)}\,^{\mu}e_{(2)}\,^{\nu}F_{\mu\nu}\,.
\end{equation}
Therefore, as $e_{a}\,^{\mu}=\delta_{a}^{\mu}$ in this frame,
\begin{equation}\label{eq33}
F_{\mu\nu}
=\left(
\begin{array}{cccc}0&0&E_{y}/c&0\\
0&0&0&0\\
-E_{y}/c&0&0&0\\
0&0&0&0
\end{array}
\right)\,.
\end{equation}
In another frame denoted as $\tilde{S}$, which is moving with a constant velocity $v$ relative to frame $S$ along the $x$ direction, the electric field can be determined through Lorentz transformations. Specifically, the components of the Faraday tensor in frame $\tilde{S}$ can be calculated as
\begin{align}
\tilde{F}_{ab}&=\Lambda_{a}\,^{(0)}\Lambda_{b}\,^{(2)}F_{(0)(2)}\nonumber\\
&=\gamma\,\left(
\begin{array}{cccc}0&0&\,E_{y}/c&0\\
0&0&\, v \, E_{y}/c^{2}&0\\
-\,E_{y}/c&\, v \, -E_{y}/c^{2}&0&0\\
0&0&0&0
\end{array}
\right)\,.\nonumber
\end{align}
In this frame, the observer measures an electric field $\tilde{E}_{y} = \gamma E_{y}$ and also observes a magnetic field $\tilde{B}_{z} = -\frac{vE_{y}}{c^{2}}$. It is important to note that the same coordinate system is employed in both frames, meaning that the transformation of the electromagnetic field between frames does not alter the coordinates. This transformation of frames serves to modify the observer's ``physical perception" of the quantities, as one observer measures only an electric field, while another measures both electric and magnetic fields, showcasing the impact of the observer's relative motion on their observations.

Distinctly, we can transform the Faraday tensor between coordinate systems without affecting the frame. Let us consider two different sets of spatial coordinates: the usual Cartesian coordinates, denoted as $x^{\mu} = (ct, x, y, z)$, and the typical cylindrical coordinates, denoted as $x'^{\mu} = (ct, \rho, \phi, z)$, where $x = \rho \cos{\phi}$ and $y = \rho \sin{\phi}$. The Faraday tensor represented in the $x^{\mu}$ coordinates can be converted to the $x'^{\mu}$ coordinates using the transformation rule defined in equation (\ref{eq8}). The resulting transformed tensor is given by
\begin{align}
\hspace{-0.5cm}F'_{\mu\nu}&=\frac{\partial x^{0}}{\partial x'^{\mu}}\frac{\partial x^{2}}{\partial x'^{\nu}}\nonumber\\
&=\left(
\begin{array}{cccc}0&\sin{\phi}E_{y}/c&\rho\cos{\phi}E_{y}/c&0\\
-\sin{\phi}E_{y}&0&0&0\\
-\rho\cos{\phi}E_{y}/c&0&0&0\\
0&0&0&0
\end{array}
\right)\,.\label{eq35}
\end{align}
In the above example, we performed a coordinate transformation that did not involve changing the frame. In the new coordinate system, the components $E_{\rho} = c\,F'^{01}$ and $E_{\phi} = c\,F'^{02}$ are non-zero, but the ``physical perception" remains unaltered, i.e., there is still only an electric field present.

\subsection{Maxwell's equations}

The dynamics of the electromagnetic field, in the presence of a source, can be determined using the variational principle. Specifically, this approach yields the form of the Faraday tensor given a four-current.

The Lagrangian of a physical system is an invariant function of the system's quantities. Out of the Faraday tensor (\ref{eq30}) and its dual (\ref{eq31}), we can construct two invariants: the scalar $F^{\mu\nu}F_{\mu\nu}$ and the pseudo-scalar $\mathcal{F}^{\mu\nu}F_{\mu\nu}$. Hence, the most simple Lagrangian densities we can construct are of second order in the Faraday tensor. If we choose the Lagrangian densities
\begin{align}
\mathcal{L}_{1}&=\frac{1}{4}k_{1}F^{\mu\nu}F_{\mu\nu}+k_{2}\,j^{\mu}A_{\mu}\,,\label{eq36}\\
\mathcal{L}_{2}&=\mathcal{F}^{\mu\nu}F_{\mu\nu}\,,\label{eq37}
\end{align}
we obtain Maxwell's theory. Notice that the source must be incorporated into the Lagrangian in a way that couples it with the field. The most straightforward coupling is through the scalar $j^{\mu}A_{\mu}$, which must be combined with another scalar. Another possibility is to consider the invariant $A^{\mu}A_{\mu}$, but this would result in a different theory, known as the Proca theory \cite{diez2020maxwell}.

By varying the Lagrangian density (\ref{eq36}) with respect to $A_{\mu}$, disregarding the surface term and applying Hamilton's principle, we obtain the tensorial field equation
\begin{equation}\label{eq38}
\nabla_{\mu}F^{\mu\nu}=\mu_{0}\,j^{\nu}\,,
\end{equation}
where $\mu_{0}\equiv k_{2}/k_{1}$.
Applying the same process to the Lagrangian density (\ref{eq37}), we obtain
\begin{equation}\label{eq39}
\epsilon_{\mu\nu}\,^{\alpha\beta}\nabla_{\alpha}F^{\mu\nu}=0\,.
\end{equation}
This set of equations (\ref{eq38}) and (\ref{eq39}) constitutes the field equations of Maxwell's theory, commonly referred to as Maxwell's equations.

In both equations (\ref{eq38}) and (\ref{eq39}), we have only one free index, with the others being summation indices. Ergo, the validity of Maxwell's equations, as measured by an observer, can be obtained by projecting the free index into a frame. Let us turn our attention to equation (\ref{eq38}). Projecting it into a frame, we obtain
\begin{equation}\label{eq40}
e^{a}\,_{\nu}\nabla_{\mu}\big( F^{\mu\nu} \big)=\mu_{0}\,e^{a}\,_{\nu}j^{\nu}\Rightarrow \nabla_{\mu}F^{\mu a}=\mu_{0}\,j^{a}\,,
\end{equation}

\noindent where we assumed $\nabla_{\mu}e^{a}\,_{\mu}=0$\footnote[4]{The covariant derivative of the tetrads is a mathematical concept that involves not only a connection related to the spacetime index but also a connection associated with the transformations of the basis, known as the spin connection denoted by $\omega_{\mu},^{ab}$. The detailed exploration of these connections and their properties is a topic that falls outside the scope of this article. The inclusion of these connections would require a more in-depth discussion of differential geometry.}. In another inertial frame, the equations can be obtained by performing a Lorentz transformation on both sides, i.e.,
\begin{equation}\label{eq41}
\Lambda^{a}\,_{b}\nabla_{\mu}F^{\mu b}=\mu_{0}\,\Lambda^{a}\,_{b}j^{b}\Rightarrow \nabla_{\mu}\tilde{F}^{\mu a}=\mu_{0}\,\tilde{j}^{a}\,.
\end{equation}

By comparing equations (\ref{eq40}) and (\ref{eq41}), we observe that Maxwell's equations transform covariantly under frame transformations. This means that if one inertial observer determines Maxwell's equations to be valid, then all inertial observers will reach the same conclusion. This covariant transformation eliminates the need for a privileged frame in the context of ME. It is important to note that in demonstrating the covariance of Maxwell's equations, we did not resort to a coordinate transformation. Additionally, both observers $S$ and $\tilde{S}$ verify the validity of the motion equations within the same coordinate system.

When writing Maxwell's equations as (\ref{eq38}) and (\ref{eq39}), we didn't specify the coordinate system because it is not necessary to demonstrate the covariance of these field equations. However, we can choose to work in a specific coordinate system if we wish. For example, in Cartesian coordinates, where $\mathring{\Gamma}^{\lambda}\,_{\mu\nu}=0$, we have
\begin{equation}\label{eq42}
\partial_{\mu}F^{\mu\nu}=\mu_{0}\,j^{\nu}\,.
\end{equation}
Alternatively, in cylindrical coordinates, we can evaluate the metric tensor in these coordinates using (\ref{eq8}), resulting in
\begin{equation}\label{eq43}
g'_{\mu\nu}=\frac{\partial x^{\alpha}}{\partial x'^{\mu}}\frac{\partial x^{\beta}}{\partial x'^{\nu}}\,g_{\alpha\beta}=\left(
\begin{array}{cccc}1&0&0&0\\
0&-1&0&0\\
0&0&-\rho^{2}&0\\
0&0&0&-1
\end{array}
\right)\,,
\end{equation}
where $g_{\alpha\beta}=diag(1,-1,-1,-1)$. From (\ref{eq17}) and (\ref{eq43}), we can evaluate the non-zero components of $\mathring{\Gamma}'^{\lambda}\,_{\mu\nu}$ as
\begin{align}
\mathring{\Gamma}'^{1}\,_{22}&=-\rho\,,\label{eq44}\\
\mathring{\Gamma}'^{1}\,_{12}&=1/\rho\,\label{eq45}\,.
\end{align}
Using definition (\ref{eq14}) and the results from equations (\ref{eq44}) and (\ref{eq45}), we can express Maxwell's equations in cylindrical spatial coordinates as
\begin{equation}\label{eq46}
\partial_{\mu}F'^{\mu\nu}+\frac{1}{\rho}F'^{2\nu}=\mu_{0}\,j'^{\nu}\,,
\end{equation}
where we used the fact that $\mathring{\Gamma}^{\nu}\,_{\mu\lambda}F'^{\mu\lambda}=0$.
In both cases, (\ref{eq42}) and (\ref{eq46}), we did not specify the frame. We can project both equations on any frame.

Equations (\ref{eq40}) and (\ref{eq41}) represent the field equations in two distinct frames, while equations (\ref{eq42}) and (\ref{eq46}) represent them in two distinct coordinate systems. These different forms and transformations are independent of each other, illustrating the versatility and independence of frame transformations and coordinate system choices in describing physical phenomena.

\subsection{Energy-momentum tensor}\label{emtensor}

When dealing with a physical field such as the electromagnetic field, it is essential to recognize that the field possesses attributes like energy, momentum, and angular momentum that can be transferred to or from particles. To work effectively with a field, it is convenient to define an energy density, as the field is distributed throughout space. This energy (and momentum) of the field can be calculated by performing a spatial integration over the region of interest. In other words, we can evaluate the local energy of an electromagnetic field within a specific region, and also determine its total global energy by integrating over the entire three-dimensional space.

The energy density and momentum density of a field are encapsulated within a rank-2 tensor known as the energy-momentum tensor. The energy-momentum tensor is defined by its conservation along a three-dimensional surface with respect to time. Consider the action for the pure electromagnetic field, i.e., without a source, given by
\begin{equation}\label{eq47}
S=\int e \,\mathcal{L}_{1} dx^{4}\,,
\end{equation}
where $e$ is the determinant of the tetrad field. We can write the determinant of the tetrads as $e=\sqrt{-g}$, where $g$ is the determinant of the metric tensor. Since the action (\ref{eq47}) is a scalar, it remains unchanged when we perform coordinate transformations, i.e., $\delta S/\delta e^{a\mu}=0$. In equation (\ref{eq47}), two quantities contribute to the variation: the tetrads $e_{a\mu}$ and the 4-potential $A^{\mu}$. The variation with respect to the 4-potential is zero, a consequence of the equations of motion. Therefore, our concern lies in the variation with respect to the tetrads.  By varying (\ref{eq47}) with respect to $e_{a\mu}$, we obtain
\begin{equation}\label{eq48}
\delta S=\int \Bigg( \frac{\partial (e\,\mathcal{L}_{1})}{\partial e_{a\mu}}-\partial_{\nu}\frac{(\partial e\,\mathcal{L}_{1})}{\partial (\partial_{\nu} e_{a\mu})}  \Bigg)\,\delta e_{a\mu} dx^{4}\,,
\end{equation}
where we have integrated by parts and dismissed the surface term. We may define the quantity
\begin{equation}\label{eq49}
\frac{1}{2}\,e\,T^{a\mu}\equiv\frac{\partial (e\,\mathcal{L}_{1})}{\partial e_{a\mu}}-\partial_{\nu}\frac{(\partial e\,\mathcal{L}_{1})}{\partial (\partial_{\nu} e_{a\mu})}
\end{equation}
and rewrite (\ref{eq48}) as
\begin{equation}\label{eq50}
\delta S=\frac{1}{2}\int e\,T^{a\mu}\delta e_{a\mu}  dx^{4}\,.
\end{equation}
By writing $T^{a\mu}\delta e_{a\mu}=\frac{1}{2}\Big( T^{a\mu}\delta_{a\mu} + T^{a\nu}\delta_{a\nu}\Big)$, we can easily show that $T^{a\mu}\delta e_{a\mu}=\frac{1}{2}T^{\mu\nu}\delta e_{\mu\nu}$ if we assume the tensor in (\ref{eq49}) is symmetrical, i.e., $T^{\mu\nu}=T^{\nu\mu}$. It is possible to demonstrate (see \S 94 of \cite{landau2013classical}) that if we assume $\delta S=0$, then
\begin{equation}\nonumber
\int e\,T^{\mu\nu}\delta g_{\mu\nu}  dx^{4}=0\,.
\end{equation}
This implies $\nabla_{\nu}(e\,T^{\mu\nu})=0$. Ergo,
\begin{equation}\label{eq52}
\nabla_{\nu}(e\,T^{a\nu})=0\,,
\end{equation}
where, again, we have used the fact that $\nabla_{\nu}e_{a\mu}=0$.
We can integrate (\ref{eq52}) over an arbitrary three-volume $V$ as
\begin{equation}\label{eq53}
\int_{V}\nabla_{\nu}(e\,T^{a\nu})dV=\int_{V}\nabla_{0}(e\,T^{a0})dV+\int_{V}\nabla_{i}(e\,T^{ai})dV=0\,.
\end{equation}
Using the divergence theorem to the term containing $\nabla_{i}T^{ai}$, we may write (\ref{eq53}) as
\begin{equation}\label{eq54}
\nabla_{0}\int_{V}e\,T^{a0}dV=-\oint dS_{i}e\,T^{ai}dV\,,
\end{equation}
where $dS_{i}$ is the surface element orthogonal to the $i$ direction of the closed surface encompassing volume $V$. By choosing the surface term at infinity, the RHS can be made to vanish and we arrive at a quantity
\begin{equation}\label{eq55}
P^{a}=\int_{V}e\,T^{a0}dV
\end{equation}
that is conserved in time. In physics, we identify a 4-vector conserved in time as the energy-momentum 4-vector of a physical system.
The quantity referred to in equation \eqref{eq49} represents the energy-momentum tensor. Within this tensor, the component $T^{(0)0}$ corresponds to the energy density, $T^{(i)0}$ represents the momentum density in the $i$-th direction, and $T^{(i)j}$ describes the elements of the stress-energy tensor. We note that the energy-momentum tensor can exhibit spacetime and frame dependency. As a result, it may assume distinct forms in various coordinate systems and frames. However, the energy-momentum 4-vector, as a real physical quantity that is measurable by an observer, must be frame-dependent but not coordinate-dependent. It should yield consistent results across different frames as long as the frames are related by Lorentz transformations of constant coefficients.

We can obtain the energy-momentum tensor for an arbitrary electromagnetic field using definition (\ref{eq49}) and the Lagrangian density $\mathcal{L}_{1}$.  For a pure field, without a source, we can rewrite (\ref{eq36}) as
\begin{equation}\label{eq56}
\mathcal{L}_{1}=\frac{1}{4}k_{1}\,g^{\mu\alpha}g^{\nu\beta}e_{a\alpha}e_{b\beta}e_{c\mu}e_{d\nu}F^{ab}F^{cd}\,.
\end{equation}
By varying (\ref{eq56}) with respect to $e_{f\lambda}$, we obtain
\begin{equation}\label{eq57}
\frac{\delta\mathcal{L}_{1}}{\delta e_{f\lambda}}=k_{1}\,F^{f\nu}F^{\lambda}\,_{\nu}\,.
\end{equation}
Thus, for the energy-momentum tensor, we have
\begin{align}
\frac{1}{2}e\,T^{a\mu}&=\frac{\delta e}{\delta e_{a\mu}}\mathcal{L}_{1}+e\frac{\delta\mathcal{L}_{1}}{\delta e_{a\mu}}\nonumber\\
&=\frac{1}{4}k_{1}e\,e^{a\mu}F^{\alpha\beta}F_{\alpha\beta}+e\,k_{1}F^{a\alpha}F^{\mu}\,_{\alpha}\,.
\end{align}
where we have used Jacob's formula for the co-factor $\delta e =e e^{a\mu}\delta e_{a\mu}$. The energy-momentum can then be written as
\begin{equation}\label{eq58}
T^{a\mu}=2k_{1}\big( e^{a\mu}F^{\alpha\beta}F_{\alpha\beta}-4e^{a}\,_{\beta}F^{\mu}\,_{\alpha}F^{\alpha\beta} \big)\,.
\end{equation}
This tensor represents the distribution of energy and momentum in an electromagnetic field.

Considering a static frame $e^{a}\,_{\mu}=\delta^{a}_{\mu}$, and using the electromagnetic field (\ref{eq33}), we can evaluate the energy $\mathcal{E}$, contained within a volume $V$. This is given by the zeroth component of the 4-momentum (\ref{eq54}) multiplied by the speed of light, i.e.,
\begin{align}\label{eq59}
\mathcal{E}&=c\,P^{(0)}=2k_{1}\int_{V}\big( e^{(0)0}F^{\alpha\beta}F_{\alpha\beta}-4e^{(0)}\,_{\beta}F^{0}\,_{\alpha}F^{\alpha\beta} \big)d^{3}x\nonumber\\
&=\frac{k_{1}}{c^{2}}\int_{V} E_{y}^{2}\,.
\end{align}
The components $P^{(i)}$ are all zero in this frame.
The energy measured by the static observer, as given in (\ref{eq59}), describes the energy of an electromagnetic field in Cartesian coordinates. If we were to describe it in cylindrical spatial coordinates, using the Faraday tensor (\ref{eq35}), we would find that the result is the same. Therefore, the choice of mathematical coordinates does not affect the measurement of energy.

Now, let us consider a different physical situation with another observer moving along the $x$ axis relative to the the static observer with constant a velocity $v$. Since we wrote the frame index in (\ref{eq58}) using the tetrads, we can easily perform a Lorentz transformation on the index $a$ of the 4-momentum, resulting in
\begin{equation}\label{eq60}
\tilde{\mathcal{E}}=c\,\tilde{P}^{(0)}=c\,\Lambda^{(0)}\,_{b}P^{b}=\gamma\,c\,\mathcal{E}\,.
\end{equation}
This means that the energy measured by the moving observer $\tilde{\mathcal{E}}$ is related to the energy measured by the static observer $\mathcal{E}$ by the Lorentz factor $\gamma$, as expressed in (\ref{eq60}). In this case, the component $\tilde{P}^{(1)} = \Lambda^{(1)}\,_{(0)}P^{(0)}$ is a non-zero quantity.
The same result (\ref{eq60}) can be obtained if we consider a tetrad field adapted to the tilde observer. In this case, there is no need to perform a Lorentz transformation, as discussed at the end of section \ref{frames}.
It is important to note that while a coordinate transformation cannot change the value of the energy density, nor can it generate a non-zero momentum for the field, a frame transformation can lead not only to a different energy measurement but also to the existence of a non-zero momentum for the field. This highlights the crucial role of frames in physics, in contrast to the lack of physical significance of mathematical coordinates, a point similarly demonstrated in Maxwell's equations.

In summary, an electromagnetic field possesses an energy density, which can be mathematically represented by its energy-momentum tensor in a specific coordinate system. When an observer measures the energy contained in a particular region of space, they project one index into a frame and perform a volumetric integral in that region. As a result, they obtain an energy-momentum $P^{a}$ that remains independent of the chosen coordinates. This emphasizes the significance of frames in physics and the lack of physical significance of mathematical coordinates when considering the fundamental properties of physical fields.

\section{Conclusions}\label{conclusions}

In this article, we have explored two fundamental concepts: one of a physical nature (frames) and another of a mathematical nature (coordinates). Both concepts play a crucial role in the description of physical systems using tensors. Tensors are geometric entities representing physical quantities that remain invariant regardless of the chosen coordinate system. Consequently, changing coordinates should not alter the underlying physical phenomena. It's important to note that when we refer to a physical phenomenon, it is inherently tied to an observer who measures it. Observers themselves are physical entities, and distinct observers may perceive different equations of motion, particularly if one of them is not in an inertial frame. This physical reality of the observer, in contrast to the mathematical nature of coordinates, implies that when changing frames, there is a physical significance associated with that frame transformation. Building upon these fundamental principles, we have presented methods for transforming both coordinates and frames.

In order to illustrate the differentiation between frame and coordinate transformations, we turned to a concrete example of a relativistic theory in section \ref{maxwell}, i.e., Maxwell's electrodynamics. Through its tensorial formalism, we elucidated how to disentangle the mathematical facet of the theory, denoted by the spacetime indices, from its physical aspect, denoted by the frame indices. We examined Maxwell's equations within two distinct coordinate systems within the same frame and also within two distinct frames using the same coordinate system. Thus, we conclusively demonstrated the independence of frame and coordinate transformations.

Tetrads represent entities that hold greater fundamental significance than the metric tensor, as we can derive the latter from the former. With 16 independent indices, tetrads encapsulate information about both the spacetime where physical phenomena unfold and the characteristics of the observer measuring these phenomena \cite{maluf2007reference}. Due to their fundamental nature, we can derive physical quantities from tetrads that would otherwise be obtained from the metric tensor, exemplified by our derivation of the energy-momentum tensor in subsection \ref{emtensor}. In this derivation, we calculated the energy-momentum tensor by varying the electromagnetic Lagrangian with respect to the tetrads, departing from the conventional approach based on the metric tensor. This yielded a tensor with one index projected into a frame, facilitating the natural definition of the 4-momentum (\ref{eq54}). The inclusion of a frame index in the 4-momentum renders the energy independent of the choice of coordinates. Had we used $P^{\mu}$ instead, we would have introduced a physical quantity contingent on coordinate choice, which would not accurately represent the empirical energy of the field. When obtaining the energy-momentum tensor through the metric tensor procedure, we are compelled to project one of the indices into a frame. However, by considering tetrads as the fundamental geometric entity, we can readily derive a genuine physical quantity without the need for additional steps.

The natural implication of the energy-momentum tensor (\ref{eq49}) is that an observer measures not an $T^{00}$ or $T^{(0)(0)}$ energy density, but rather an $T^{(0)0}$ energy density. This can have significant implications for our interpretation of natural phenomena. In nature, we consistently observe that the energy density of known sources (not limited to electromagnetic ones) is always positive. Theoretical sources that violate this almost empirical principle are termed exotic matter. If we interpret the measured energy density as $T^{00}$, we may enforce the condition $T^{00}>0$. However, if the energy density is understood as $T^{(0)0}$, the condition $T^{00}<0$ does not necessarily imply $T^{(0)0}<0$, as we must specify the observer making the measurement. There are physical scenarios in which we have $T^{00}<0$, but $T^{(0)0}>0$ for real physical observers. One example is the energy density $T^{(0)0}$ of the hypothetical source responsible for the Alcubierre spacetime \cite{alcubierre1994warp}, which yields positive values for stationary observers and negative values for unphysical ones capable of traveling faster than light \cite{carneiro2022total}.

The utilization of tetrad fields as fundamental variables leads to possibilities in the construction of gravitational theories where these fields serve as the primary variables. One such theory is the Teleparallel Equivalent to General Relativity (TEGR), which consistently yields fundamental quantities. For instance, TEGR provides a 4-momentum for the gravitational field, resulting in an energy expression akin to that presented in (\ref{eq54}) \cite{maluf2013teleparallel}. Within TEGR, there are several analogies between the gravitational and electromagnetic fields when an observer undergoes a boost, for example, for Schwarzschild geometry, the gravitoelectromagnetic components transform in a similar way, in the linear regime, to the fields of Maxwell's theory \cite{gravitomagnetismo}.

The formulation of electromagnetic quantities with tetrad-projected indices enables a coordinate-independent description of Maxwell's theory. These features may be significant in the theory for next-generation detectors of gravitational waves (GWs). Current laser interferometers are limited to detecting low-frequency GWs, and new detectors based on the emission of electromagnetic waves in the presence of GWs may overcome these limitations \cite{wolfram}.

Throughout this article, our focus has been primarily on inertial observers. However, it is worth noting that we can extend our analysis to include accelerated observers. When doing so, we encounter another connection related to frame indices, known as the Levi-Civita connection, which arises when calculating the covariant derivative. This connection assumes a fundamental role, particularly when demonstrating the relation $\nabla_{\nu}e^{a}\,_{\mu}=0$ that we have utilized. For comprehensive details regarding this extension, readers are encouraged to refer to Ref. \cite{maluf2010electrodynamics}, as it falls beyond the scope of the present article.

It is also pertinent to note an intriguing parallel that warrants further exploration. In quantum mechanics, we know that the choice of basis in which a state is expressed does not alter the quantum state itself, but rather dictates how measurements are conducted and interpreted. For instance, consider polarized light represented as a quantum superposition of vertical and horizontal polarization states. When we measure polarization in this basis (vertical/horizontal), we effectively choose to observe the light in these specific terms. However, if one wishes to observe circular polarization, the basis must be changed to one that describes right and left circular polarization states. This change in basis alters how we interpret the measurement results but does not modify the actual physical state of the light. In essence, the choice of basis determines the ``lens'' through which we observe and measure the quantum system, but it does not change the underlying quantum state. This concept is pivotal in quantum mechanics and reflects the theory's probabilistic and non-deterministic nature. The analogy between choosing a reference frame in General Relativity and selecting a basis for a quantum state in Dirac notation offers a compelling perspective on the interplay between spacetime geometry and quantum states. This parallelism, though not the focus of our current study, suggests a fertile ground for future research, particularly in understanding the quantum properties that might influence spacetime structures, such as those in the Alcubierre warp drive \cite{carneiro2022total,alcubierre1994warp}. Such explorations could potentially bridge gaps in our understanding of quantum mechanics and general relativity, contributing to the broader quest for a unified theory in physics.


\end{document}